\RequirePackage{amsmath}
\documentclass{llncs}
\usepackage{times}
\usepackage{graphicx}
\usepackage{epsf}
\usepackage{verbatim}
\usepackage{float}
\usepackage{amsfonts}
\usepackage{amssymb}
\usepackage{ulem}
\usepackage{lmodern}
\usepackage{hyperref}
\usepackage{xcolor}
\usepackage{comment}

\newcommand{\nmedio}{\overline{n}}
\newcommand{\wmedio}{\overline{w}}
\newcommand{\nbarra}{\underline{n}}

\begin{document}

\title{Computational Algorithms for the Product Form Solution of Closed Queuing Networks with Finite Buffers and Skip-Over Policy}
\author{Gianfranco Balbo$^{1}$, Andrea Marin$^{2}$, Diletta Olliaro$^{2}$, Matteo Sereno$^{1}$}
\institute{$^1$$\,$Dipartimento di Informatica, Universit\`{a} di Torino, Turin, Italy\\
$^2$$\,$Dipartimento di Informatica, Universit\`{a} Ca' Foscari Venezia, Venice, Italy}

\maketitle
\begin{abstract}
    Closed queuing networks with finite capacity buffers and skip-over policies are fundamental models in the performance evaluation of computer and communication systems. This technical report presents the details of computational algorithms to derive the key performance metrics for such networks. The primary focus is on the efficient computation of the normalization constant, which is critical for determining the steady-state probabilities of the network states under investigation. A convolution algorithm is proposed, which paves the way for the computation of key performance indices, such as queue length distribution and throughput, accommodating the intricacies introduced by finite capacity constraints and skip-over mechanisms. Finally, an extension of the traditional Mean Value Analysis algorithm addressing numerical stability is provided. The approaches discussed here allow make the investigation of large-scale networks feasible and enable the development of robust implementations of these techniques for practical use. 
\end{abstract}

\section{Introduction}

Queuing networks with finite capacity buffers and blocking mechanism have been find very useful in the modelling of computer and communication systems and production and manufacturing systems as well. 
Various blocking mechanisms have been analyzed in the literature to represent the behaviours of system with limited resources. A comprehensive surveys of the results concerning these models is presented in~\cite{Balsamo:2011} where the very rich list of references can be used to study the details of most of the work published in the literature on this subject.

When certain conditions are met, the underlying model of a queuing network with blocking is a Markov Chain which admits a product form solution (see~\cite{Balsamo:2001}). While this property usually implies the possibility of computing standard performance indices in a computationally efficient manner, this is not always the case for these type of networks where additional constraints are introduced either on the characteristics of the service function~\cite{Balsamo:1998}, or on the topology of the network~\cite{Clo:1998}, or on the policy adopted when the server is blocked~\cite{Sereno:1999}.

In this paper, we consider closed queuing networks where each server has a finite capacity buffer and where jobs that join saturate queues (i.e., queues whose input buffer is ful) skip such queues in search for another queue with non-full input buffer.
These networks have been proposed by Pittel~\cite{pitt79}and have been proved to admit a product form solution~\cite{pitt79,MRO23}.

Taking into account the peculiarities and intricacies posed by the presence of finite capacity buffers and of the skip-over policies introduced to specify the behavior of a customer attempting to join a queue with a full buffer, we specialize some of the algorithms published in the literature for the efficient solution of product form queuing networks.

When the state spaces of the product form models are finite, the boundary conditions introduce dependencies that make the computation of the performance metrics of interest computationally complex, except for the cases in which the solutions can be expressed in a recursive manner. Blocking mechanisms and skip-over policies make these recursions more difficult to obtain resulting, however, in algorithms that are only marginally more complex of their counterparts developed for BCMP queueing networks.

The paper presents detailed derivations of these new algorithms that are kept simpler restricting the attention to single class models (models where all the customers are statistically identical), leaving as a future work the burden of extending the notation and the derivations to multiple class models which are often difficult to obtain because of the complexity of the notation and not because of additional conceptual complications.

The paper is organized as follows. Section~\ref{sec:defandnormconst} gives a formal definition of the queuing networks under investigation and presents a convolution algorithm for the computation of the normalization constant. Section~\ref{sec:perfindices} provides the explicit expressions for the computation of several key performance metrics. Section~\ref{sec:mva} extends the traditional Mean Value Analysis algorithm to accommodate the unique characteristics of these networks with blocking. Section~\ref{sec:conclusions} concludes this work. Finally, the appendices provide additional details on the techniques discussed, addressing various computational challenges.

\section{Skip-Over Queuing Networks and Normalisation Constant}\label{sec:defandnormconst}

Consider a closed queuing network with $M$ service stations and $N$ customers. Each station $i$ ($i = 1,2,...,M$) has an input buffer which can accommodate $C_i$ customers (included that in service). Moreover, all the stations have constant service rates $\mu_i = 1/S_i$, where $S_i$ is the expected value of the negative-exponential distribution of the service time.\\
The interconnections among the stations are captured by a routing matrix $\bf Q$.
Denoting with $X_i(n)$ the throughput of station $i$ when there are $n$ customers in the network ($n = 1, 2, ..., N$), the routing matrix allows to write
\begin{eqnarray}\label{Eq:XB}
    \mathbf X(n) \ = \ \mathbf X(n) \, \mathbf Q
\end{eqnarray}
The routing matrix has rank $M-1$, thus admitting an infinite number of parallel solutions.
Choosing one of the stations as the reference one (say station $1$), define $V_i$ as the average number of visits that a customer makes to station $i$ between two passages from the reference station. From this definition follows that $V_1 = 1$ and that the above indefinite system of equations can be transformed in definite one whose solution is represented by the vector $\bf V$ that collects all these visit ratios and whose expression is the following
\begin{eqnarray}\label{Eq:FB}
    \mathbf V \ = \ \mathbf V \, \mathbf Q
\end{eqnarray}

From the above equation, it is possible to show that for any pair of stations $i$ and $j$  (independently of $n$)
\begin{eqnarray}\label{Eq:Visits}
    \displaystyle{\frac{X_i(n)}{X_j(n)}} \ = \ \displaystyle{\frac{V_i}{V_j}}
\end{eqnarray}

Networks of this type have been shown \cite{pitt79,MRO23,OBMS23} to admit a product form expression for the state probability which can the be written in the following way \\
\begin{eqnarray}\label{EQ.:PFS}
P(\nbarra) \ =\ \frac{1}{G} \prod_{i=1}^{M} f_i(n_i)
\end{eqnarray}
where $n_i$ denotes the number of customers at station $i$, $G$ is a normalization constant  and
\begin{eqnarray}\label{Eq:sf}
f_i(k)= \left\{
\begin{array}{l l}
1 & k = 0\\
V_iS_i f_i(k-1) & 0 < k \leq C_i\\
0 & C_i < k\\
\end{array}
\right.
\end{eqnarray}
is an auxiliary function called {\em {service function of station $i$}} defined in terms of quantities specific of such station only.\\

In a closed system of this type, a state
$
\nbarra = ( n_1, n_2, ..., n_M)
$
is defined as the distribution of customers over the stations of the network and
is {\bf feasible} if\\[-4mm]
\begin{eqnarray}
0 \leq n_i \leq C_i, i = 1,2,.,M
~~~~~~{\mbox {and} }~~~~~~
\sum_{i=1}^M {n_i} = N
\end{eqnarray}
where the second condition introduces  the dependency  among the stations of the network.\\[1mm]

For notational convenience, define the state space of a system with $m$ stations and $n$ customers in the following manner\\[-5mm]

\begin{eqnarray}
S(n,m, {\bf c}) =
\left\{
(n_1, n_2, ..., n_m)\
|\
0 \leq n_i \leq c_i, \ i=1, 2, ..., m;\ \
\sum_{i=1}^m {n_i} = n
\right\}
\end{eqnarray}
where $\bf c$ is a vector, called {\em {capacity vector}} with $m$ components.

Notice that the presence of these capacity limits restricts the possible value of the total number of customers circulating in the network ($n$) which cannot exceeds the limit
\begin{eqnarray}\label{Eq:nmax}
n \ \leq \ n_{max} \ = \ \sum_{i=1}^{m} c_i
\end{eqnarray}

\begin{figure}[t]
\centering
\includegraphics[width=0.5\linewidth]{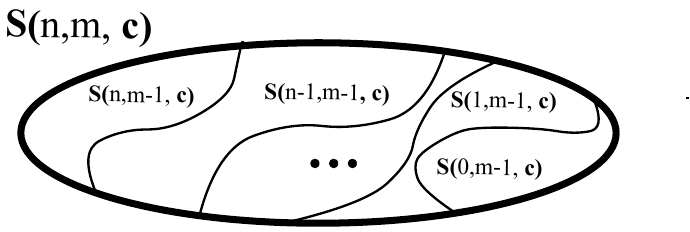}
\caption{Partition of a Closed Pittel Network State Space}
\vspace{-15mm}
\label{fig:partitionStateSpacePittel}
\end{figure}

Following the notation and the approach introduced in \cite{buze73,denn78,ARBaBr83b}, let us define the following auxiliary function
\begin{eqnarray}
g(n,m,{\bf c}) &=&
\sum_{\nbarra \in S(n,m, {\bf c})} {\ \prod_{i=1}^m {f_i(n_i)}}
\end{eqnarray}
 such that
$g(N,M, {\bf {C}}) = G$ (the normalization constant of EQ.~\ref{EQ.:PFS}).
With a little manipulation of this definition (that we will show next), it is possible to show that
\begin{equation}\label{Eq:gnm}
\begin{split}
g(n,m,{\bf c}) &=
\sum_{k=0}^{n} {f_m(k) g(n-k, m-1, {\bf c})}\\
&=
\sum_{k=0}^{L_m} {f_m(k) g(n-k, m-1, {\bf c})}
\end{split}
\end{equation}
where $L_m = min(n,c_m)$ and the summation stops at $L_m$ because of the fact that the auxiliary function $f_m(k) = 0$ for $k > c_m$ (as said in \ref{Eq:sf}).

This result comes from a  derivation which relies on the possible partition of the state space of the network into proper sub-spaces characterized by the number of customers at the $m$-th station (from left to right, $0$, $1$, ... $n$) as shown in the following figure.\\
Indeed, we can write
\begin{equation}
\begin{split}
g(n,m, {\bf c}) & =  {\sum_{\nbarra \in S(n,m, {\bf c})} {\prod_{i=1}^m
{f_i(n_i)}}}\\[-2mm]
&= {\sum_{k=0}^{n} \left\{ \sum_{\stackrel{\nbarra \in S(n,m, {\bf c})} {n_m=k}}
\left[{\prod_{i=1}^{m}
{f_i(n_i)}}\right]\right\}}\\[-2mm]
&=\sum_{k=0}^{n} f_m(k) \left\{ \sum_{\nbarra \in S(n-k,m-1,{\bf c})}
\left[{\prod_{i=1}^{m-1}
{f_i(n_i)}}\right]\right\}\\[-2mm]
&=\sum_{k=0}^{L_m} f_m(k) g(n-k,m-1, {\bf c})
\end{split}
\end{equation}
where the last step of this derivation accounts for the truncated expression of $f_m(k)$

The definition of $g(n,m, {\bf c})$ can be made recursive if we introduce the following initial conditions\\[-0mm]
$g(0,m, {\bf c}) = 1 \ (m = 0,1,.)$ and $g(n,0, {\bf c}) = 0\
(n = 1,2....)$

The recursive definition of the auxiliary function is a convolution sum which gives the name {\it Convolution Algorithms} to the techniques developed to compute the normalization constant in an efficient manner.

 A careful analysis of the way this recursion  works, allows to notice that for the computation of $g(n,m, {\bf c})$, we must have first computed all the values of $f_m(k)$ and of $g(k,m-1, {\bf c}), k = 0,1,...,n$.

Notice that the computation of $g(n,m, {\bf c})$ is quite simple when the stations have a {\it Load Independent} behavior, since in this case the auxiliary functions $f_i(k)$ are powers of the basic quantity $Y_i = V_iS_i$, for $i=0,1,..., c_m$ and then are set to $0$ for higher values of $k$.
\begin{equation}
\begin{split}
g(n,m, {\bf c}) &\!\! =\!\! \sum_{k=0}^{n} {f_m(k) g(n\!-\!k,m\!-\!1, {\bf c})}\\[-0.5mm]
&\!\! =\!\!g(n,m\!-\!1, {\bf c}) + \sum_{k=1}^{n} {f_m(k) g(n\!-\!k,m\!-\!1, {\bf c})}\\[-0.5mm]
&\!\! =\!\!g(n,m\!-\!1, {\bf c}) + \sum_{k=1}^{L_m} {(Y_m)^k g(n\!-\!k,m\!-\!1, {\bf c})}\\[-0.5mm]
&\!\! =\!\!g(n,m\!-\!1, {\bf c}) + Y_m \sum_{k=1}^{L_m} {(Y_m)^{k\!-\!1} g((n\!-\!1)\!-\!(k\!-\!1),m\!-\!1, {\bf c})}\\[-0.5mm]
&\!\! =\!\!g(n,m\!-\!1, {\bf c}) + Y_m \sum_{h=0}^{L_m-1} {(Y_m)^{h} g((n\!-\!1)\!-\! h,m\!-\!1, {\bf c})}
\end{split}
\end{equation}

The presence of the limit $L_m-1$ in the last line of the previous derivation does not allow to recognize the summation as the definition of $g(n-1,m, {\bf c})$ except when $n \leq c_m$.

Indeed, it is worth to recall the explicit expression of $g(n-1,m, {\bf c})$ to highlight the term that is missing in the last line of the previous derivation, i.e., the extension of the summation to $L_m$ (or $c_m$ in this case of interest).

\begin{eqnarray}
g(n-1,m, {\bf c}) = \left\{
\begin{array}{l l}
1 & \ \ n = 0\\
\ \\
\displaystyle{\sum_{k=0}^{n - 1} (Y_m)^{k} g((n\!-\!1)\!-\!k,m\!-\!1, {\bf c})} & \ \  n-1 \leq c_m\\
\ \\
\displaystyle{\sum_{k=0}^{c_m} (Y_m)^{k} g((n\!-\!1)\!-\!k,m\!-\!1, {\bf c})} & \ \ c_m < n-1 < n_{max}\\
\end{array}
\right.
\end{eqnarray}

We can thus derive the following expression for the computation of the normalization constant
\begin{eqnarray}
g(n,m, {\bf c}) = \left\{
\begin{array}{l l}
1 & \ \ n = 0\\
\ \\
g(n,m\!-\!1, {\bf c}) + Y_m \ g(n\!-\!1,m, {\bf c}) & \ \ 0 < n \leq c_m\\
\ \\
\displaystyle{g(n,m\!-\!1, {\bf c}) + Y_m \ \sum_{h=0}^{c_m - 1} (Y_m)^{h} g((n\!-\!1)\!-\!h,m\!-\!1, {\bf c})} & \ \ c_m < n \leq n_{max}\\
\end{array}
\right.
\end{eqnarray}

To clarify this issue, and for later reference, it is also possible to provide an alternative form for this last expression, which is however not well suited for implementation purposes

\begin{eqnarray}
g(n,m, {\bf c}) = \left\{
\begin{array}{l l}
1 & \ \ n = 0\\
\ \\
g(n,m\!-\!1, {\bf c}) + Y_m \ g(n\!-\!1,m, {\bf c}) & \ \ 0 < n \leq c_m\\
\ \\
g(n,m\!-\!1, {\bf c}) + Y_m \ g(n\!-\!1,m, {\bf c}) - (Y_m)^{c_m} g((n\!-\!1)\!-\!c_m,m\!-\!1, {\bf c}) & \ \ c_m < n \leq n_{max}\\
\end{array}
\right.
\end{eqnarray}

Denoting with $C_{max}$ the size of the largest buffer of the stations,  we can conclude that the complexity of this computation method  (expressed in terms of the number of multiplications needed for obtaining $g(N,M, {\bf C})$) is of the order of $C_{max}*N*M$

\section{Performance Indices}\label{sec:perfindices}

The computation of the normalization constant yields intermediate values that are useful to compute some performance measures also in the case of these networks with blocking.\\

\subsection{Queue Length Distribution}
Consider first the computation of the queue length distribution of the $m-$th station
\begin{eqnarray}
 p_m(k,n, {\bf c}) =
 \sum_{\genfrac{}{}{0pt}{}{ \nbarra \in S(n,m, {\bf c})}{n_m = k}} P(\nbarra)
 \end{eqnarray}
Using the explicit expression of $P(\nbarra)$ we have\\[-2mm]
\begin{equation}
\begin{split}
 p_m(k,n, {\bf c}) &=
 \frac {f_m(k)}{g(n,m, {\bf c})}
 \sum_{\nbarra \in S(n-k,m-1, {\bf c})}
 \prod_{j=1}^{m-1} {f_j(n_j)}\\
 & =
 f_m(k) \frac {g(n-k,m-1, {\bf c})} {g(n,m, {\bf c})}
\end{split}
\end{equation}
 with the understanding that when we write $\nbarra \in S(.,m-1, {\bf c})$,
 we refer to a capacity vector ${\bf c}$ of $m-1$ components. Moreover, we must realize that when $n$ is larger than $c_m$, the components of the distribution are equal to zero for all the values of $k$ larger than $c_m$.\\

 The queue length distribution for the other stations of the network cannot be obtained without some additional computation.  Owing to the commutative property of the convolution operator, the normalization constant of a Product Form Queuing network is independent of the order in which the different stations are considered.

 A common practice used to compute the queue length distribution of an arbitrary station is that of
 reordering the stations in such a way that the selected station appears as the last one  and then repeating the whole computation from the beginning.
 In Appendix A we will present an alternative method to obtain the desired result that becomes computationally convenient when the network consists of more than $3$ stations.\\

 For the time being, let us define the normalization constant computed for the network in which the $i-th$ station is missing (or better, is {\em {short-circuited}}, also just {\em {shorted}} for the sake of brevity) as
 \begin{eqnarray}
g^{[-i]}(n,m,{\bf c}) &=&
\sum_{\stackrel {\nbarra \in S(n,m, {\bf c})}{n_i=0}} {\ \prod_{\stackrel{j=1}{j \neq i}}^m {f_j(n_j)}}
\end{eqnarray}

Then, it is easy to show that the queue length distribution of the $i$-th station can be computed using the following expression
 \begin{equation}\label{Eq:QLD-i}
 \begin{aligned}
 p_i(k,n, {\bf c}) &=
 \frac {f_i(k)}{g(n,m, {\bf c})}
 \sum_{\stackrel{\nbarra \in S(n-k,m, {\bf c})}{n_i=0}}
 \prod_{\stackrel{j=1}{j \neq i}}^{m} {f_j(n_j)}\\
 & =
 f_i(k) \ \frac {g^{[-i]}(n-k,m, {\bf c})} {g(n,m, {\bf c})}
\end{aligned}
\end{equation}

\subsection{Throughput}

The knowledge of the queue length distributions, allows to easily compute the throughputs of the stations of the network.\\

The presence of limited buffers requires however that we discuss first the impact that they have on the formal definition of the throughput in our networks.

Let us focus on the behaviour of the $m-th$ station.  The flow of customers leaving the station may results from the mixing of two different components. Whenever the number of customers in the network exceeds the size of the buffer of the $m$-th station, some of the customers depart from the station after receiving service from the server, while others leave the station without having received any service because when they arrived at the station they  found the input buffer full and thus skipped the station.  Let us denote with $X_m^{[P]}(n)$ the (Productive) throughput of customers that received the service and with $X_m^{[S]}(n)$ that of customers that Skipped the station. $X_m(n) = X_m^{[P]}(n) + X_m^{[S]}(n)$ is then the total throughput of the station.

The throughput of customers that received service from the station is naturally computed using the expression of the queue length distribution and is thus defined in the following manner
\begin{equation}
\begin{split}
 X_m^{[P]}(n) &=
 \sum_{k=1}^n {p_m(k,n, {\bf c}) \frac {1} {S_m}}\\
 &= \frac {1} {S_m}\  \sum_{k=1}^{L_m} {p_m(k,n, {\bf c})}
\end{split}
\end{equation}
Using the explicit expression of
$p_m(k,n, {\bf c})$ and recalling the definition of $f_m(k$), we derive the following result

 \begin{equation}
 \begin{split}
 X_m^{[P]}(n) & =  \frac {1} {S_m} \
 \sum_{k=1}^{L_m} {V_mS_m f_m(k-1)
 \frac {g(n-k,m-1, {\bf c})} {g(n,m, {\bf c})} }\\
 & =  {V_m} \
 \sum_{k=1}^{L_m} {f_m(k-1)
 \frac {g(n-k,m-1, {\bf c})} {g(n,m, {\bf c})} }\\
  & =  \frac{V_m}{g(n,m, {\bf c})} \
 \sum_{k=1}^{L_m} {(Y_m)^{k-1}
 {g(n-k,m-1, {\bf c})}} \\
  & =  \frac{V_m}{g(n,m, {\bf c})} \
 \sum_{h=0}^{L_m-1} {(Y_m)^h
 {g((n-1)-h,m-1, {\bf c})}}
\end{split}
\end{equation}

With a reasoning similar to that used to derive the final expression of $g(n,m)$ and of $p_i(k,n,{\bf c})$, we can obtain the following computationally efficient expression that is valid independently of the reference station chosen for the computation of the $\{ V_i\}$

\begin{eqnarray}\label{Eq:ProdX}
X_i^{[P]}(n) = \left\{
\begin{array}{l l}
0 & \ \ n = 0\\
\ \\
\displaystyle {V_i \  \frac {g(n-1,m, {\bf c})} {g(n,m, {\bf c})}} & \ \ 0 < n \leq c_i\\
\ \\
\displaystyle {\frac{V_i}{g(n,m)} \
 \sum_{h=0}^{c_i-1} {(Y_i)^h
 {g^{[-i]}(n-k,m, {\bf c})}}}  & \ \ c_i < n\\
\end{array}
\right.
\end{eqnarray}



The two expressions for the Productive throughput provided by this last equation allow to make a few important observations.

The flow equations that are used to compute the visit counts are written under the hypothesis of balancing the throughput in and out of a service station independently of the fact that it comes from customers that received a service or customers that skipped the station. When writing these equations, we consider a station as a box including the server, the buffer and the skipping path. We can thus conclude that the Total throughput defined before as $X_i(n)$ is a component of the solution vector of the traffic equations so that as soon as we know the value of one of these components we can obtain the value of all the others by using the visit counts (see Eqs.~\ref{Eq:FB} and \ref{Eq:Visits}).

When $n \leq c_i(n)$, no customer will ever find the buffer of station $i$ full, making the Productive and Total throughputs of the $i-th$ station  identical. The expression of $X_i^{{P}} (n)$ from  Eq.~\ref{Eq:ProdX} when $n \leq c_i$ provides a value that satisfies the flow equations and that can thus be used to identify the Total throughputs of all the stations of the network (notice that, whenever $n$ is both  $\leq c_j$ and $\leq c_k$, Eq.~\ref{Eq:ProdX}, evaluated for $i=j$ and $i=k$, gives values that agree with this observation).\\

When $n$ is larger that the buffer sizes of all the stations of the network, the situation is much less clear since station-skipping can now occur for all the stations and we are no longer in the position of identifying in a simple manner the components of the Total throughput vector, since we would need to know at least one of them to compute the others. The conceptual difficulty of this problem is made evident if we consider the case in which the network is loaded with a number of customers $n$ identical to $n_{max}$ (defined in Eq.~\ref{Eq:nmax}).
Only one feasible state exists, while all the stations are working at full speed (the Productive Throughput of each station is identical to its nominal rate) and a substantial additional flow of customers among the stations derives from the intensive skipping activity that is taking place in this situation (every customer that leaves a specific station at the completion of its service ends up to join the same station, which has now one position free in its buffer, after attempting to move to several other stations whose buffers are all full). The vector of the Productive throughput does not satisfy the flow equations and a careful investigation of this situation is needed in order to identify which of the many solutions of the flow balance equations is the right one.\\

Referring again to Eq.~\ref{Eq:ProdX}, we can notice that the difference between the second and the third lines of the equation is due to the fact that the summation on the third line stops at $c_i -1$ instead of $c_i$.
Indeed, we can notice that extending the summation to $c_i$, i.e., adding one element in the summation, we could recognize such summation  as the definition $g(n-1,m, {\bf c})$.,

We can thus {\bf{conjecture}} that this additional element is the skipping component of the total flow which then assumes the following form

\begin{eqnarray}\label{Eq:conjecture}
X_i^{[S]}(n) \ =\
\left\{
\begin{array}{l l}
0 & \ \ n \leq c_i\\
\ \\

\displaystyle {V_i  \, (Y_i)^{c_i}\  \frac {g^{[-i]}(n-1-c_m,m, {\bf c})} {g(n,m, {\bf c})}} & \ \ c_i < n \leq N_{max}
\end{array}
\right.
\end{eqnarray}
 On the basis of this conjecture, that we discuss at some length in Appendix C,  we can claim that the total throughput of the $i$-th station has the following expression, valid for any (feasible) value of the population of the network.

\begin{eqnarray}\label{Eq:TotThrough}
X_i(n) \ =\ \displaystyle {V_i \,  \frac {g(n-1,m, {\bf c})} {g(n,m, {\bf c})}} & \ \ 0 < n \leq N_{max}
\end{eqnarray}

\subsection{Utilization}

The utilization of the
 $m-$th station derives immediately from the knowledge of the queue length distribution

 \begin{eqnarray}
 U_m(n) = 1 - p_m(0,n) = 1 - \frac {g(n,m\!-\!1)} {g(n,m)}
 \end{eqnarray}

 and can be obtained efficiently from the computation of the normalization constant.

Using arguments similar to those discussed for the derivation of the recursive expression of $g(n,m)$ we can state  that
 \begin{eqnarray}
U_m(n) = \left\{
\begin{array}{l l}

X_m(n) S_m & \ \ 0 < n \leq C_m\\
\ \\
 \displaystyle{1 - \frac {g(n,m-1)} {g(n,m)}} & \ \ C_m < n
\end{array}
\right.
\end{eqnarray}

The expression of the utilization of the $m$-th station can be generalized to that of the $i$-th station using the expression of the queue length distribution derived before (see Eq.~\ref{Eq:QLD-i}) thus obtaining
 \begin{eqnarray}
U_i(n) = \left\{
\begin{array}{l l}

X_i(n) S_i & \ \ 0 < n \leq C_i\\
\ \\
 \displaystyle{1 - \frac {g^{[-i]}(n,m)} {g(n,m)}} & \ \ C_i < n
\end{array}
\right.
\end{eqnarray}

\subsection{Mean Queue Length and Mean Waiting Time}

 The computation of the mean queue length can only be obtained using its definition (the standard formula) and thus with additional costs
\begin{eqnarray}
 \nmedio_i(n)= \sum_{k=1}^{n} {k p_i(k,n)}
\end{eqnarray}
When we have $\nmedio_i$, we use Little's formula to derive the mean waiting time

 \begin{eqnarray}
 \wmedio_i = \frac{\nmedio_i(n)}{X_i(n)}
\end{eqnarray}
where this last expression accounts for the customers that are not waiting at the station when they skip it.

\section{Mean Value Analysis}\label{sec:mva}

The extension of Mean Value Analysis to networks with limited buffers requires first to derive a new recursive formula for the queue length distribution.\\
Exploiting the recursive expression of the service function $f_m(k)$, we can write

%
\begin{equation}
\begin{split}
p_m(k,n) & =
 Y_m \ f_m(k-1)\  \frac {g(n-k,m-1)} {g(n,m)}\
 \frac{g(n-1,m)} {g(n-1,m)}\\
& =  V_m \ \frac {g(n-1,m)} {g(n,m)}\  S_m f_m(k-1)\
 \frac {g(n-k,m-1)} {g(n-1,m)}\\
& =  X_m(n) S_m \ p_m(k-1, n-1)
\end{split}
\end{equation}
since the $m$-th station cannot accommodate more than $c_m$ customers, we have
\begin{eqnarray}
p_m(k,n) &  = \ \left \{
\begin{array}{l l}
\displaystyle {X_m(n) S_m \ p_m(k-1, n-1)]} & \ \  k \leq c_m
\ \\
0 & \ \ k > c_m
\end{array}
\right.
\end{eqnarray}

Given that  the expression of the Total throughput $X_m$ holds true also for any index $i$ (with proper adjustment), we get
\begin{eqnarray}\label{Eq:QLD-FIRST}
p_i(k,n) &  = \ \left \{
\begin{array}{l l}
\displaystyle {X_i(n) S_i \ p_i(k-1, n-1)]} & \ \  k \leq c_m
\ \\
0 & \ \ k > c_m
\end{array}
\right.
\end{eqnarray}
This recursive expression for the queue length distribution, together with Little's formula  allow to derive a convenient expression for the mean waiting time at the $i-th$ station.
\begin{eqnarray}
{\overline w}_i(n) = \frac{{\overline n}_i(n)}{X_i(n)} = \frac{\displaystyle{\sum_{k=1}^{L_i} {k
\ p_i(k,n)}}} {X_i(n)}
\end{eqnarray}
Introducing the expression of the queue length distribution we get
\begin{equation}
\begin{split}
{\overline w}_i(n) & =  \frac{\displaystyle{\sum_{k=1}^{L_i} {k \ X_i(n) S_i \
p_i(k-1,n-1)}}} {X_i(n)}\\
& =  \sum_{k=1}^{L_i} {k S_i \ p_i(k-1,n-1)}
\end{split}
\end{equation}

The recursive definition of the total throughput obtained from the average waiting times using Little's formula allows also the derivation of an interesting recursive expression for the Skipping throughput.
Indeed, referring to Eq.~\ref{Eq:conjecture}, we have
\begin{equation}
\begin{split}
X_i^{[S]}(n) \ &= \ \displaystyle {V_i \, (Y_i)^{c_i} \, \frac {g(n-1-c_i,m, {\bf c})} {g(n,m, {\bf c})}}
\\
&=\displaystyle {V_i \  (Y_i)^{c_i} \, \frac {g(n-1,m, {\bf c})} {g(n-1,m, {\bf c})}}\
\displaystyle {\frac {g(n-1,m, {\bf c})} {g(n,m, {\bf c})}}\\
&=\displaystyle {V_i \, p_i(c_i,n-1, {\bf c})}\
\displaystyle {\frac {g(n-1,m, {\bf c})} {g(n,m, {\bf c})}}\\
&=\displaystyle {X_i(n) \  p_i(c_i,n-1, {\bf c})}
\end{split}
\end{equation}

 Summarizing, we have the following recursive equations for the MVA algorithm derived for this type of networks
\begin{equation}\label{Eq:MVA-Al}
\begin{split}
{\overline w}_i(n) & =  \sum_{k=1}^{L_i} {k S_i \ p_i(k-1,n-1)}\\
X_{ref}(n) & =  \ \frac {n} {\displaystyle{\sum_{j=1}^M {V_j {\overline w}_j(n)}}}\\
X_i(n) & =  V_i X_{ref}(n)\\
\nmedio_i(n) & =   {\wmedio_i(n)} {X_i(n)}\\
p_i(k,n) & =  X_i(n) S_i\  p_i(k-1, n-1)\ \ \ k=1,2, ..., n\\
U_i(n) & =   \sum_{k=1}^{L_i} p_i(k,n)\\
p_i(0,n)  & =  1.0 - U_i(n)\\
X_i^{[S]}(n) \ &=
0 \ \ \ \ \ \ \ \ \ \ \ \ \  \ \ \ \ \ \ \ \ \ \ \ \ \ \ \ \ n \leq C_i\\
\ & = \displaystyle {X_i(n) \  p_i(c_i,n-1, {\bf c})} \ \ \ \  n > C_i\\
X_i^{[P]}(n) \ &=\displaystyle {\frac {1}{S_i} U_i(n)}
\end{split}
\end{equation}

The algorithm is initialized, setting for all the stations of the network
\begin{eqnarray}
p_i(0,0) = 1.0 \ \ \ \ i=1,2, ..., M
\end{eqnarray}

Notice that the algorithm can be conveniently simplified when $n \leq C_i,  i = 1, 2, ..., M$. Indeed, in this case
\begin{equation}
\begin{split}
{\overline w}_i(n) & =  \sum_{k=1}^{C_i} {k S_i \ p_i(k-1,n-1)}\\
& =  \sum_{k=1}^{n} {k S_i \ p_i(k-1,n-1)}\\
& =  S_i \sum_{k=1}^{n} {(1 + (k-1))  \ p_i(k-1,n-1)}\\
& =  S_i \left[ \sum_{h=0}^{n-1} { \ p_i(k-1,n-1)} +
\sum_{h=0}^{n-1} {h  \ p_i(h,n-1)}
\right]\\
& =  S_i \left[ 1.0 +  \nmedio_i(n-1)
\right]\\
\end{split}
\end{equation}

from which we also derive the simpler expressions for
\begin{eqnarray}
U_i(n) = X_i(n) S_i
\end{eqnarray}
and
\begin{eqnarray}
\nmedio_i(n) = U_i(n) \left[ 1.0 +  \nmedio_i(n-1)
\right ]
\end{eqnarray}

Notice however that, if we want to compute the performance indices of the stations for $n$ larger than the minimum of all the buffer sizes, the computation of the queue length distributions for all the stations of th e network must be carried on  since both, the general expression for ${\overline w}_i(n)$ and  $X_i^{[S]}(n)$ (and as a consequence ${\overline n}_i(n)$) depend on it as we can see from the following formulas
\begin{eqnarray}
{\overline w}_i(n) \ = \ \left \{
\begin{array}{l l}

\displaystyle {S_i\,\left[ 1.0 + \nmedio_i(n-1) \right]}
 & \ \ 0 < n \leq c_i\\
\ \\
\displaystyle {\sum_{k=1}^{L_i} {k S_i \ p_i(k-1,n-1)}} & \ \ c_i < n
\end{array}
\right.
\end{eqnarray}
and
\begin{eqnarray}
X_i^{[S]}(n)\ = \ \left \{
\begin{array}{l l}
0  & \ \ 0 < n \leq c_i\\
\ \\
\displaystyle {X_i(n) \  p_i(c_i,n-1, {\bf c})} \ \ \ \  n > c_i
\end{array}
\right.
\end{eqnarray}

The computation of $p_i(0,n)$ is formally correct, but may yield numerical instability problems when the $i$-th station is heavily utilized.

We can thus conclude that this new version of the MVA algorithm, suited for closed queuing networks with load independent stations and limited buffer sizes is formally correct at the expenses of
\begin{itemize}
\item a computational cost that is larger than that for closed queuing network with load independent stations and unlimited buffers,
\item  a possible numerical instability problem that becomes evident when the load $N$ of the network becomes closer to the sum of the buffer sizes of all the networks.
\end{itemize}

The numerical instability problems that affect this new version of the MVA algorithm, extended to address the peculiar features of the networks discussed in this paper, has already been analyzed in the discussion of the standard Mean Vale Analysis algorithm applied to network with Load Dependent servers~\cite{reis80,reiser:mvald}. In Appendix D we recall the basic results presented originally in \cite{reiser:mvald} and we provide a modified version of our extended MVA algorithm that overcomes these numerical instability problems that affect our networks even in the case of Load Independent servers.

\section{Conclusions}\label{sec:conclusions}

The implementation of the results derived in this paper requires to address complexity and numerical stability issues that are peculiar of the Convolution and Mean Value Analysis methods when extended to the solution of Closed Queuing Networks with Finite Buffers and Skip-Over Policies. The techniques discussed in Appendices A and D address these computational problems and allow to derive robust implementations that can be conveniently used in practice.

With these algorithms it is possible to derive and to study the behaviours of the performance indices under variable load conditions ($N$) showing, for instance, how the queue length distributions of the different stations of the network evolve when $N$ becomes larger and larger, up to the point of exceeding the sum of the buffer sizes of all the stations of the network.

All the results derived in this paper have been implemented and tested solving a large set of non-trivial models.

\bibliographystyle{IEEEtran}
\bibliography{biblio}

\section{Appendix A}

A common approach for the computation of the queue length distribution of the station of index $i$ different from $M$ is that of repeating the computation from the beginning re-indexing the stations of the network so that the station for which we want to compute all its performance indices appears as the last considered by the algorithm for the computation of the normalization constant. Because of this, it is commonly assumed that the computational complexity for the computation of the performance indices of all the stations of the network is $M$ times that of the computation of the performance indices of the station of index $M$.
A computationally efficient algorithm, called {\em{inverse convolution method}}, exists to obtain these desired results (see, for instance \cite{bruell-balbo:convolution}) which is however numerically unstable and thus rarely used in practice.\\

Trading time with space, it is possible to design an algorithm that computes the performance indices of all the stations of the network with a computational complexity that is of the order of $3$ times that of the computation of the indices of the last station without introducing any numerical instability problem.
This method exploits the commutative property of the convolution operator which implies that the normalization constant of a product form queuing network is independent of the order in which the different stations are considered.\\

Given an arbitrary ordering of the $M$ stations of the network, we can define the following two auxiliary functions $g_{UP}(n,m,{\bf{c}})$ and $g_{DW}(n,m,{\bf{c}})$ that are simple variations of the definition introduced by Eq.~\ref{Eq:gnm}
\begin{equation}\label{Eq:gnmUP}
\begin{split}
g_{UP}(n,m,{\bf c}) &=
\sum_{k=0}^{L_m} {f_m(k) \ g_{UP}(n-k, m-1, {\bf c})}\ \ \ \ m=1,2,...,M
\end{split}
\end{equation}
with $g_{UP}(n,1,{\bf{C}}) = f_1(n) \ \ \ \ n = 0,1,..., N$ and
\begin{equation}\label{Eq:gnmDW}
\begin{split}
g_{DW}(n,m,{\bf c}) &=
\sum_{k=0}^{L_m} {f_m(k) \ g_{DW}(n-k, m+1, {\bf c})}\ \ \ \ m=M-1, M-2, ...1
\end{split}
\end{equation}
with $g_{DW}(n,M,{\bf{C}}) = f_M(n) \ \ \ \ n = 0,1,..., N$.\\

The recursive computation of these two sets of quantities allows to derive directly the queue length distribution of the station of index $M$

 \begin{equation}\label{Eq:QLD-M}
 \begin{split}
 p_M(k,N, {\bf C}) &=
 f_M(k) \ \frac {g_{UP}(N-k,M-1, {\bf C})} {g_{UP}(N,M, {\bf C})}
\end{split}
\end{equation}

and of that of index $1$

 \begin{equation}\label{Eq:QLD-1}
 \begin{split}
 p_1(k,N, {\bf C}) &=
 f_1(k) \ \frac {g_{DW}(N-k,2, {\bf C})}
  {g_{DW}(N,1, {\bf C})}
\end{split}
\end{equation}

The above quantities  can also be used to derive with a single convolution step the values of $g^{[-i]}(N,M,{\bf{C}})$ in the following manner

\begin{equation}\label{Eq:gnmDW-i}
\begin{split}
g^{[-i]}(N,M,{\bf {C}}) &=
\sum_{k=0}^{N} {g_{UP}(k,i-1,{\bf{C}})} \ {g_{DW}(N-k, i+1, {\bf {C}})}\ \ \ \ i=2,3,..., M-1
\end{split}
\end{equation}

which makes possible to derive the queue length distribution of the $i$-th station in the usual manner
\begin{equation}\label{Eq:QLD-ibis}
\begin{split}
 p_i(k,N, {\bf C}) &=
 f_i(k) \ \frac {g^{[-i]}(N-k,M, {\bf c})} {g(N,M, {\bf C})}
\end{split}
\end{equation}
where $g(N,M, {\bf {C}})$ is either equal to  $g_{UP}(N,M, {\bf {C}})$ or $g_{DW}(N,1, {\bf {C}})$.

The algorithm implemented using these equations is robust, numerically stable, and capable of performing correctly even in the extreme cases of queuing networks where all the stations have buffers of limited sizes and the number of customers in the network is either equal to the sum of all these dimensions (in this case there is only one possible state) or exceeds this sum (thus making the situation unfeasible).

\section{Appendix B}
\subsection{The Flow Equivalent Server}

The analysis of the behaviour of a Product Form queuing network when the parameters of one station may vary can be conveniently performed using the method of the Flow Equivalent Server (FES) proposed originally by Chandy, Herzog, and Woo \cite{chw}.

Assume that the station whose parameters may change is the last of the network (the station of index $m$), then a composite station which captures the behaviour of the first $m-1$ stations may be preliminarily constructed so that a set of two station models (models comprising the FES and the $m$-th station with different service times) can be solved in a computationally convenient manner to obtain the behaviour of the whole system under these varying conditions. The method was called originally the "Norton's theorem for queuing networks" (\cite{chw}), because of its similarity with the Norton's theorem commonly used for the analysis of electrical circuits.

The method consists in solving the original network with the $m$-th station removed (or shorted i.e., with its average service time $S_m$ set to $0$) for different values of the number of customers $k=1,2,...,n$ computing the throughput that flows from the subnetwork toward the $m$-th station, and using these throughput values as the service rates of the load dependent FES which characterizes the composite behaviour of the "constant" part of the network.

\begin{figure}[ht]
\centering
\includegraphics[width=0.5\linewidth]{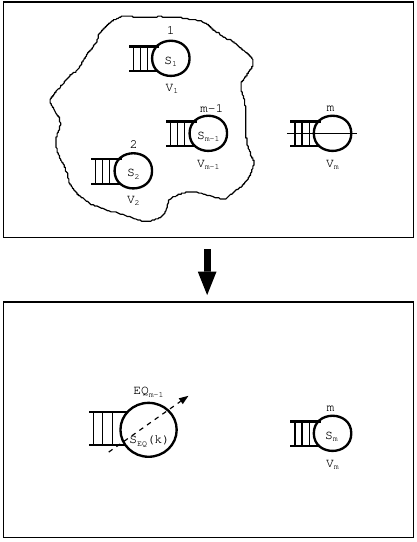}
\caption{}
\label{fig:flowEqNortonT}
\vspace{-15mm}
\end{figure}

From a computational point of view, this corresponds to calculate the normalization constants of the sub-network comprising the first $m-1$ stations for different level of load $k=1,2,   ,N$ and use them to compute te desired throughput (this procedure is often called a "Controlled Experiment", see \cite{denn78,ARBaBr83b}). According to the results reported in \cite{ARBaBr83b}, and using the notation introduced at the beginning of this paper, we can define
\begin{equation}\label{Eq:VEQ}
\begin{split}
V_{EQ} \ &= \displaystyle{ \sum_{i=1}^{m-1} V_j \ q_{i,m} } \ =\ Vm
\end{split}
\end{equation}
which is in agreement with the formula used in the standard convolution algorithm to compute the throughput of an arbitrary station of the network.
The load dependent average service time of the Equivalent Server is thus
\begin{equation}\label{Eq:FES}
\begin{split}
S_{EQ}(k) \ &= \displaystyle{ \frac{1}{X_{EQ}(k)}} \ =\ \frac{1}{V_{EQ}} \ \frac{g(k,m-1,{\bf{c}}}{g(k-1,m-1,{\bf{c}}} \ \ \ k=1,2, N
\end{split}
\end{equation}
The solution of the two station model (i.e., the computation of the normalization constants of the composite model, from which all the performance measures of interest can be obtained) thus involves  the computation of the service function of the FES and that of the $m$-th station which depends on the parameters of such station for the different cases of the analysis.
Because of the construction, the visit counts of the two stations of the composite models are both equal to $V_m$;
then, from the above definition of the service rate of the composite server, we write

\begin{eqnarray}
f_{EQ}(k) \ = \left\{
\begin{array}{l l}
1   &  k\ =\ 0\\
V_m \ {\displaystyle {\frac{1}{V_m}}} \ {\displaystyle {\frac{g(k,m-1,{\bf{c}}}{g(k-1,m-1,{\bf{c}}} \ f_{EQ} (k-1)}} & k\ =\ 1,2, n
\end{array}
\right .
\end{eqnarray}

Making this recursion explicit,  we have
\begin{equation*}
\left\{
\begin{aligned}
f_{EQ}(0) & \,=\, 1\\
f_{EQ}(1) & \,=\, V_m \, f_{EQ}(0) \, \frac{g(1,m\!-\!\!1,{\bf{c}})}{ V_m \ g(0,m\!-\!\!1,{\bf{c}})}   \,=\, g(1,m\!-\!\!1,{\bf{c}})\\
f_{EQ}(2) &  \,=\, V_m \,  f_{EQ}(1)\,  \frac{g(2,m\!-\!\!1,\bf{c})}{V_m \  g(1,m\!-\!\!1,\bf{c})}   \,=\, V_m \ g(1,m\!-\!\!1,{\bf{c}}) \ \frac{g(2,m\!-\!\!1,{\bf{c}})}{V_m \ g(1,m\!-\!\!1,{\bf{c}})}   \,=\, g(2,m\!-\!\!1,{\bf{c}})\\
f_{EQ}(3) &  \,=\, V_m \, f_{EQ}(2) \,   \frac{g(3,m\!-\!\!1,\bf{c})}{V_m \ g(2,m\!-\!\!1,\bf{c})}   \,=\, V_m \ g(2,m\!-\!\!1,{\bf{c}})\  \frac{g(3,m\!-\!\!1,{\bf{c}})}{V_m \ g(2,m\!-\!\!1,{\bf{c}})}  \,=\, g(3,m\!-\!\!1,{\bf{c}})\\
...\\
f_{EQ}(k) & \,=\, V_m \, f_{EQ}(k\!-\!\!1) \, \frac{g(k,m\!-\!\!1,{\bf{c}})}{V_m \ g(k\!-\!\!1,m\!-\!\!1,{\bf{c}})}   \,=\, V_m \ g(k\!-\!\!1,m\!-\!\!1,{\bf{c}})\  \frac{g(k,m\!-\!\!1,{\bf{c}})}{V_m \ g(k\!-\!\!1,m\!-\!\!1,{\bf{c}})}   \,=\, g(k,m\!-\!\!1,{\bf{c}})\ \ \  k \leq C_{max}^{[m\!-\!\!1]}\\
f_{EQ}(k) &  \,=\, 0\ \ \  k > C_{max}^{[m\!-\!\!1]}
\end{aligned}
\right.
\end{equation*}
thus concluding that
\[
f_{EQ}(k) \ =\   g(k,m-1,{\bf{c}}) \ \ \ \ \  k = 1, 2, ...,  C_{max}^{[m\!-\!\!1]}
\]
The recursive structure of the convolution algorithm characterized by Eq.~\ref{Eq:gnm} shows that the computation of the normalization constant of a (sub)network defined on the first $i$ service stations depends only on the service function of the $i-th$ station ($f_i(k)$) and on the vector whose components are the normalization constants computed for the (sub)network containing the first $i-1$ stations with a variable number of customers ($g(k,i-1,\bf{c})$).

Actually, the $i$-th step of the algorithm that constructs the normalization constant for the (sub)network containing the first $i$ stations is insensitive with respect to the way in which the vector containing the normalization constants for the previous (sub)network - the (sub)network containing the first $i-1$ stations - has been computed and, specifically, on the number of convolution steps that have been performed to produce such results.

Considering that the convolution algorithm is initialized with the equation
\begin{eqnarray}
g(k,1,{\bf{c}}) \ =\  f_1(k) \ \ \ k = 0, 1, ..., c_1
\end{eqnarray}
we can generalize the reasoning that we made for the $m$-th station, stating that, after each step of the algorithm, we can construct a special station  whose service function is directly equivalent to $g(k,i-1,\bf{c})$. Its equivalent visit count is $V_{eq} \ = \ V_i$ and
its average (load dependent) service time  $S_{eq}(k)$ is
\begin{eqnarray}
S_{eq}(k) = \displaystyle{\frac {1}{V_i}\ \frac{g(k,i-1,\bf{c})}{g(k-1,i-1,\bf{c})}}
\end{eqnarray}
so that we get
\begin{eqnarray}
f_{eq}(k) = {g(k,i-1,\bf{c})}
\end{eqnarray}
The throughput of customers that flows from this subnetwork toward the $i$-th station, that is used to define $S_{eq}(k)$
\begin{eqnarray}
X_{eq}(k) = \displaystyle{{V_i}\ \frac{g(k-1,i-1,\bf{c})}{g(k,i-1,\bf{c})}}
\end{eqnarray}
can also be interpreted as the "view" that the $i$-th (shorted) station has of the behaviour of the current subnetwork.

\begin{figure}[ht]
\centering
\includegraphics[width=0.5\linewidth]{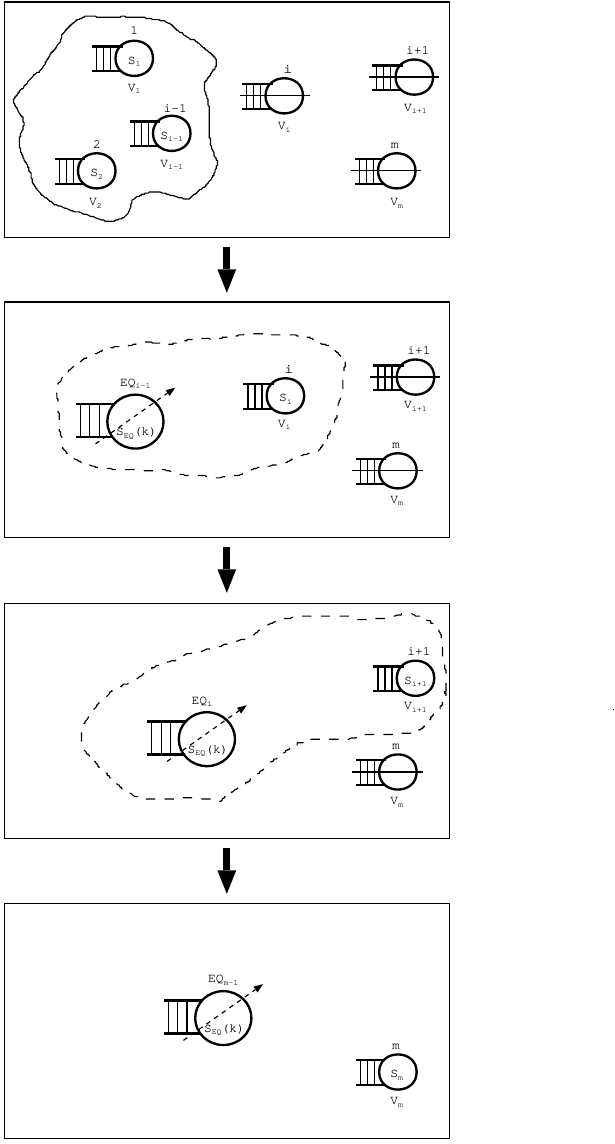}
\caption{}
\label{fig:flowEqGen}
\vspace{-15mm}
\end{figure}

\section{Appendix C}

\subsection{ Proof of the Conjecture}

To support the validity of our conjecture we must consider the following two elements
\begin{itemize}
\item Solution of the traffic equations
\item Relationships between skipping and network population\\
\end{itemize}

\subsection {Traffic equations}
\ \ \\
The routing matrix $\mathbf Q$ describes the way in which customers move from one station to the others, without considering the skipping paths. This means that the service stations of the networks considered in these notes can be envisioned as "black boxes" which include the server, the input buffer and the skipping path. The solutions of the traffic equations, which can be considered as average visit counts, provide the average number of visits that customers exiting from the reference station make to the other stations, before returning to the reference one {\em {independently}} of the fact that they are leaving the reference station either after being served or after  skipping the server due to a full buffer.
The ratios between different visit counts correspond thus to the ratios between the total throughputs out of the "black boxes".

Notice that the visit counts are computed independently of the service speeds of the different stations and of the total number of customers circulating in the network. These ratios would be the same independently of the fact that some of (or all) the buffers were infinite and that some of the service speeds were infinite too.\\

\subsection {Skipping and Total throughputs}

Given that the convolution method for the computation of the normalization constant of the product form solution is independent of the order in which the different stations are considered, let us assume that the buffer sizes of the stations of the network are all different from each-others and that the service stations are indexed so that $1$ is the index of the station with the smallest buffer and $M$ is the index of the station with the largest buffer.\\

Suppose to discuss the behavior of the network for an increasing population size.

Skipping starts to occur on the station with the smaller buffer when $n$ (the size of the population) is equal to $c_1 + 1$. For smaller values of $n$ the throughputs of the different stations are all "Productive Throughputs" whose relative values agree with the ratios of their corresponding visit counts.

When the first skipping occurs, customers who arrive at station $1$ and find its buffer full are immediately directed to one of the other stations according to the routing probabilities where they will find space in its input buffer. With this population size, no skips will occur on the other stations whose throughputs will be computed using Eq.~\ref{Eq:ProdX}. The ratios among these other throughputs will be coherent with the solution of the traffic equation and the skipping throughput of station one easily computed either as
\begin{itemize}
\item the difference between the "Total throughput" determined via the traffic equations and the "Productive Throughput" computed via the corresponding queue length distribution.\\
\item or using Eq.~\ref{Eq:conjecture} which identifies the term that is making the "Productive Throughput" of station $1$ different from its "Total Throughput".
\end{itemize}

Increasing the number if customers in the network, we reach the point in which the skipping occurs both on station $1$ and station $2$. Also in this case the "Total Throughput" of station $m$ is identical to its "Productive Throughput" and the ratios between the visit counts allow to compute the "Total Throughputs" of all the stations of the network. These total throughputs turn out to be identical to the corresponding "Global Throughputs" of all the stations , but the first two where the differences are due to the missing terms
identified by  Eq.~\ref{Eq:conjecture}
whose evaluation provides the "Skipping Throughputs" of these two stations.\\

These observations remain valid up to the point in which skipping occurs on all the stations of the network, but the last one.

When the population size becomes large enough to induce skipping on all the stations of the network, we can only observe that the "Productive Throughputs" (that we can always compute using the queue length distributions) do not satisfy the traffic equations.

The problem in this situation is that of finding a way to compute the total throughputs of all the stations which are now all subject to skipping.
The rank of the routing matrix is such that we can identify an infinite set of "parallel" solutions (solutions which differ by a multiplicative constant), among which we must identify the correct one.

The FES idea (discussed in Appendix B) can be used to prove the validity of our conjecture in this more general situation assuming that we extend the original network adding a fictitious station with an infinite buffer next to the $m$-th station and with a routing connection chosen to leave unvaried the solution of the traffic equations for what concerns the visit counts of all the stations of the original network.

Because of its unlimited buffer, and
independently of the service rate value of this station, there will never be skipping on this new station so that its Total Throughput will always coincide with the Productive Throughput defined before (see Eq.~\ref{Eq:ProdX}).\\

The  application of the Convolution Algorithm to this extended network will consist of $m+1$ steps.
When performing the $m$-th step, the algorithm computes a normalization constant vector ($g(k,m,{\bf{c}}), k=0,1,...,n$) identical to the final one of the original network.
Using the Flow Equivalent Server approach outlined before, we can thus compute the service rate of an equivalent server that interacts with the added station. The service rates of this equivalent station are computed using the formula of Eq.~\ref{Eq:TotThrough}.
The last step of the convolution algorithm (the step of index $m+1$) can be interpreted as the method for computing the normalization constant of a simple "tandem" network consisting of two stations only (the equivalent one and the added one) and thus as the basis for the solution of this reduced, equivalent model. Obviously  in this last context, the throughput of the equivalent server depends on the value of the service rate of the additional station which also defines its (total) throughput. Moreover, the traffic equations of the extended network allow in this case to define the Total Throughputs of all the stations of the original network (via the simple ratios among their visit counts).

In principle, we can progressively increase the speed of the added station observing a perturbation of the behaviour of the original network that becomes smaller and smaller. When the speed $\mu_{m+1}$ is very large, so that $S_{m+1} \approx 0$, the behaviours of the original and of the extended networks become approximately equal, while is still possible to observe that the Productive and Total troughputs of the added station are identical and the Total throughputs of all the stations of the network are thus still computable.

Now, assume that the average service time of the added station is equal to $0$, the algorithm will produce a vector $g(k,m+1, \bf{c})$ which is identical to $g(k,m,\bf{c})$. In this situation, when $k$ customers are in the network, no customers are ever queuing at the added station. This implies that they are all at the composite server that thus produces a constant output rate  identical to its  service rate evaluated at level $k$.

This is the value that we are missing.

Having found it, using the visit counts, we can directly compute the Total Throughput of the added station  and, as a consequence,   the Total throughputs of all the stations of the network that are then given by Eq.~\ref{Eq:TotThrough}, as well as the Skipping ones that result from the evaluation of
Eq.~\ref{Eq:conjecture}.

\ \\
\section{Appendix D}

The numerical instability problem that may derive from a direct implementation of the formulas contained in  Eq.~\ref{Eq:MVA-Al}, can be overcome addressing three different aspects of the problem
\begin{enumerate}
\item Finding a new recursive expression for the computation of the queue length distribution
\item Obtaining the solution of a general (product form) queuing network through the successive aggregation of stations, as already discussed in Appendix B when observing the possibility of finding the FES for the subnetworks containing the first $i$ stations
\item Updating the queue length distributions of the first $i$ stations computed by solving the tandem network in which the $i$-th station interacts with the equivalent server representing the aggregate of the first $i-1$ stations when a new aggregate server is constructed for the next step of the iteration.
\end{enumerate}

\subsection{Alternative Queue Length Distribution Formula}

The definition of the queue length distribution of Eq.~\ref{Eq:QLD-ibis} can be used as a starting point to derive an alternative recursive expression that can be used to overcome the numerical instability problem deriving from the need of defining the probability of an empty queue as the complement to its utilization. Indeed it is possible to obtain this result with a simple manipulation of the original expression as described next.

\begin{equation}\label{Eq:NEW-i}
 \begin{split}
 p_i(k,n, {\bf c}) &=
 f_i(k) \ \ \frac {g^{[-i]}(n\!-\!k,m, {\bf c})} {g(n,m, {\bf c})}\\
 &=
 f_i(k) \ \ \frac {g^{[-i]}((n\!-\!1)\!-\!(k\!-\!1),m, {\bf c})} {g(n,m, {\bf c})}\
  \ \ \frac {V_i}{V_i} \
 \ \ \frac {g^{[-i]}((n\!-\!1)\!-\!k,m, {\bf c})} {g^{[-i]}((n\!-\!1)\!-\!k,m, {\bf c})}\
 \ \ \frac {g((n\!-\!1),m, {\bf c})} {g((n\!-\!1),m, {\bf c})}\\
&=
 f_i(k) \ \ \frac {g^{[-i]}((n\!-\!1\!-\!k),m, {\bf c})} {g(n\!-\!1,m, {\bf c})}\
 \ \ \frac {V_i \ g((n\!-\!1),m, {\bf c})} {g(n,m, {\bf c})}\
 \ \ \frac {g^{[-i]}((n\!-\!1\!-\!k)\!+\!1,m, {\bf c})} {V_i \ g^{[-i]}((n\!-\!1\!-\!k),m, {\bf c})}\\
 &=
  p_i(k,n-1, {\bf c})\ \ \frac { X_i(n)} {Y_i(n\!-\!k)}
\end{split}
\end{equation}
where $Y_i(n-k) = V_i \ g^{[-i]}((n-1-k),m, {\bf c}) \ / \ g^{[-i]}((n-k), m, {\bf c})$ can be interpreted as the throughput of customers flowing through the $i$-th station when it is shorted (i.e., its service time is set to $0$) and $n-k$ customers are circulating in the rest of the network.

Considered alone, this result suffers from the same numerical instability problem of the standard MVA formula (see Eq.~\ref{Eq:QLD-FIRST}) because it is now the probability of a full queue that derives from a noirmalization argument.
However the combination of these two results allows to express the desired queue length distribution in the following manner.
\begin{eqnarray}\label{Eq:NEW-QLD}
p_i(k,n, {\bf c})  \ =\
\left\{
\begin{array}{l l}
 X_i(n) \ p_i(0,n-1, {\bf c})\ \ {\displaystyle {\frac { 1} {Y_i(n-k)}}}\  & \ \ k = 0\\
 \ \\
 X_i(n) S_i \ p_i(k-1, n-1) & \ \ 0 < k \leq n
\end{array}
\right.
\end{eqnarray}
where $n$ ranges from $0$ to $c_i$.

\subsection{ Solution of a Tandem Network}

Eq.~\ref{Eq:NEW-QLD} shows that it is possible to derive all the values of the queue length distribution with no numerical instability problems thus avoiding the execution of any subtraction that may yield cancellation errors. This is however possible at the cost of knowing  two quantities that are not usually available: the throughput of the queue at level $n$ ($X_i(n)$) which is often dependent on the queue length distribution itself and the throughput of the same service station $Y_i(n-k)$ computed when it is shorted (its service time is set to $0$) and $n - k$ customers are circulating in the rest of the network.
This difficulty can be successfully faced relaying on a solution approach that we have already mentioned when discussing the definition of a Flow Equivalent Server.

Remaining within the context of MVA, where the performance indices are computed with a recursion on the number of customers in the network, we can introduce another recursive "direction" by solving first a network containing stations $1$ and $2$, then, using this result, consider the network containing stations $1$, $2$, and $3$, up to the point in which the whole original network is analyzed. The recursion is initialized considering station $1$ in isolation so that its throughput for the different levels of load $n=1,2, ..., c_1$ is $X(n) = 1/S_1$ and the queue length distributions for the different loads are set equal to $1$ only when the number of customers in the first queue is equal to the load itself.

To continue the discussion and to make it simple, let us introduce the following notation.
\begin{itemize}
\item $p\_EQ_i(k,n)$ is the probability that $k$ customers are within the sub-network comprising the first $i$ stations, while $n$ customers are in the whole network.
\item $X\_EQ_i(k)$ is the throughput of the $i$-th station when $k$ customers are within the sub-network comprising the first $i$ stations, considered in isolation (i.e., when all the stations of index $j > i$ are shorted).
\item $C\_EQ_i$ is the sum of the buffer sizes of the first $i$ stations of the network (i.e., the maximum number of customers that can be accommodated in the $i$-th subnetwork - the subnetwork comprising the first $i$ stations).
\item $w\_EQ_i(k)$ is the time spent by the customers within the $i$-th subnetwork (i.e.,the subnetwork comprising the first $i$ stations), when $k$ customers are within such sub-network.
\item $Y\_EQ_i(k)$ is the throughput of the station of index $i+1$ when considered "in tandem" with the $i$-th subnetwork, while its service time is still set to $0$ (i.e., when the $(i+1)$-st station is still shorted), when $k$ customers are within such sub-network.
\end{itemize}

The analysis of the entire network is recursively performed solving a set of tandem networks in which the first server is the composite station equivalent to the $i$-th subnetwork (i.e., the global behaviours of the first  $i$ stations) and the second server is just station of index $i+1$. The recursion is repeated for $i=1,2, ..., m-1$ and initialized by considering station $1$ as the composite server of index $1$. As already mentioned before, the initialization consists in  setting $X\_EQ_1(k) = 1/S_1, k = 1,2, ..., c_1$ and $0$ otherwise.

The $i$-th  step of the recursion involves
\begin{enumerate}
\item computing the throughput of the $(i+1)$-st (shorted) station
\[
Y\_EQ_i(k) = X\_EQ_i(k) \ V_{i+1} / V_i), \ \ \ \ \emph{}k = 1,2, ..., N
\]
\item initializing the queue length distribution of the $(i+1)$-st station
\[
p_{i+1}(0,0) = 1.0
\]
\item increasing the load of the tandem network $n$ from $1$ to $N$, compute
\begin{itemize}
\item the waiting time in the composite server
\[
w\_EQ_i(n) = \sum_{k=1}^{min(n,C\_EQ_i)} k \ p_{i+1}(n-k,n-1)/Y\_EQ_i(k)
\]
\item the waiting time of the $(i+1)$-st server
\[
w_{i+1}(n) = \sum_{k=1}^{min(n,c_i)} k \ S_{i+1} \ p_{i+1}(k-1,n-1)
\]
\item the cycle-time $CYT[n]$ of the tandem network
\[
CYT(n) = w\_EQ_i(n) + w_{i+1}(n)
\]
\item the throughput of the composite server
\[
X\_EQ_i(n) = n / CYT(n)
\]
\item the queue length distribution of the $(i+1)$-st station\\
\item[] \ \ \ \ \ \ \ \ \ \ if $n <= C\_EQ_i$\\
\[
p_{i+1}(0,n) = p_{i+1}(0,n-1)\  X\_EQ_i(n)  /  Y\_EQ_i(n)
\]
\ \ \ \ \ \ \ \ \ \ \ \ \ \ \ \ \ \ \ \ \ \ \ \ \ \ \ \ \ \ \ and, for $k = 1, 2, ..., min(n,c_i)$,
\[
\ \ \ \ \ \ p_{i+1}(k,n, = p_{i+1}(k-1,n-1)\  X\_EQ_i(n) \ S_{i+1}
\]
\ \ \ \ \ \ \ \ \ \ otherwise\\
\[
p_{i+1}(k,n) = 0, \ \ \ \ \ k = 0, 1, ..., min(n,c_i)\ \ \ \ \ \ \ \ \ \
\]
\item the queue length distribution of the composite server
\[
p\_EQ_i(k,n) = p_{i+1}(n-k,n)\ \ \ \ \  k = 0, 1, ..., min(n,c_i)
\]
\ \ \ \ \ \ \ \ \ \ \ \ \ \ \ \ \ \ \ \ \ \ \ \ \ \ \ while the other components are set to $0$

\end{itemize}
\item noticing that the value of $X\_EQ_i(n), n = 1,2, ..., C\_EQ_i$ computed during the $i$-th step of the recursion is the basis for starting the next step.
\end{enumerate}

\subsection{ Updating the Queue Length Distributions of all the Stations}

The computation of the queue length distribution of the $i$-th composite server allows to update the queue length distribution of the stations collectively represented by such composite server performing a weighted sum.
Let $p_j^{[i-1]}(k,n)$  be the probability that $k$ customers where at the $j$-th station when $n$ customers where in the composite server comprising the first $i-1$ stations and that  $p\_EQ_{i-i}(n,n)$
is the probability of having $n$ customers (globally) in the composite server, then  for each ($j$-th) station in the composite server we can update the components of the queue length distributions in the following way
\[
p_j^{[i]}(k,n) = {\displaystyle { \sum_{l=k}^n \ p_j^{[i-1]}(k,l)\ p\_EQ_{i-i}(l,l)}}
\]
\subsection{ Computation of the Performance Indices for all the Stations}

The knowledge of the queue length distributions for all the stations of the network (and for all the possible loads), together with the throughput of the station of index $M$ allows to derive in a simple manner all the missing measures.

For what concerns the total throughputs we can easily obtain them as
\[
X_i(n) = {\displaystyle { X_m(n) \ \frac{V_i}{V_m}}}, \ \ \ i=0,1,2, ..., M\ \ ~{\mbox {and} } n=1,2, ..., N
\]

Using the expressions contained in Eq.:\ref{Eq:MVA-Al}, for all possible values of $i$ and $n$,   we have

\begin{equation*}
\begin{split}
{\overline w}_i(n) & =  \sum_{k=1}^{L_i} {k S_i \ p_i(k-1,n-1)}\\
\nmedio_i(n) & =   {\wmedio_i(n)} {X_i(n)}\\
U_i(n) & =   1.0 - p_i(0,n)\\
X_i^{[P]}(n) \ &=\displaystyle {\frac {1}{S_i} U_i(n)}\\
X_i^{[S]}(n) \ &=
0 \ \ \ \ \ \ \ \ \ \ \ \ \  \ \ \ \ \ \ \ \ \ \ \ \ \ \ \ \ n \leq C_i\\
\ & = \displaystyle {X_i(n) \  p_i(c_i,n-1, {\bf c})} \ \ \ \  n > C_i
\end{split}
\end{equation*}

\end{document}